\documentclass{elsart}

\usepackage{graphicx}

\usepackage{amssymb}

\begin{document}

\begin{frontmatter}



\title{Correlation of coming limit price with order book in stock markets}


\author[heisei,nict]{Jun-ichi Maskawa}
\ead{maskawa@heisei-u.ac.jp}
\address[heisei]{Department of Management Information, Fukuyama Heisei University, Fukuyama, Hiroshima 720-0001,
 Japan}
\address[nict]{NiCT, 2-2-2 Hikaridai Seiki-cho Soraku-gun, Kyoto 619-0288, Japan (ATR)}

\begin{abstract}
We examine the correlation of the limit price with the order book,
when a limit order comes. We analyzed the Rebuild Order Book of
Stock Exchange Electronic Trading Service, which is the
centralized order book market of London Stock Exchange. As a
result, the limit price is broadly distributed around the best
price according to a power-law, and it isn't randomly drawn from
the distribution, but has a strong correlation with the size of
cumulative unexecuted limit orders on the price. It was also found
that the limit price, on the coarse-grained price scale, tends to
gather around the price which has a large size of cumulative
unexecuted limit orders.
\end{abstract}

\begin{keyword}
Limit order \sep Order book \sep Stock market \sep Data analysis
\PACS 07.05.Kf \sep 89.65.Gh
\end{keyword}
\end{frontmatter}

\section{Introduction}
\label{intro}

The fat tail of the price fluctuation and the long memory of the
volatility are common features observed in financial markets
employing the continuous double auction as the mechanism for price
formation \cite{gopik1998,plerou1999,liu1999,cont2001}. The origin
of such features is an important problem to be solved in
econophysics and finance, and still under debated. Many models
that give rise to one of those or both features have been proposed
\cite{bouchaud2003}.

Among those models, Maslov has studied a simple model of markets
driven by continuous double auctions \cite{maslov2000}. In his
model, traders choose one from the two types of orders at random.
One is a market order to sell or buy at the best available price
at the time when the trader places the order. The other is a limit
order to sell or buy with the specification of the limit price,
which is the worst allowable price for the trader. A new limit
order to sell (buy) is placed above (below) the current market
price with a random offset drawn from a uniform distribution in a
given interval. The size of order is fixed for simplicity.
Numerical simulation has shown that the distribution of the price
fluctuation generated by this model has a power-law tail, and the
volatility has a long range correlation. Although his model has
some unsatisfactory details about the statistics of the price
fluctuation \cite{note1}, the approach of microscopic market
models to this problem is convincing, because they describe the
actual process of price formation from the ultimate microscopic
description level. Several authors have proposed the market models
in the same class \cite{challet2003,smith2003,maskawa2006}.

The way traders select a limit price is random and non-strategic
in Maslov's model. It is a simplification of the actual way of
traders to select a limit order. In the paper \cite{maskawa2006},
traders are assumed to be mimetic, and prefer the limit price
which has a large stock of limit orders, when they place limit
orders. Incorporating this tendency of coming limit price into the
microscopic model of stock market, the author reproduces the price
fluctuation with the more real statistics depending on the
parameter of the model.

In this paper, we empirically examine the correlation of the limit
price with the condition of order book at the time when a limit
order comes. We analyze the Rebuild Order Book of Stock Exchange
Electronic Trading Service (SETS), which is the centralized order
book market of London Stock Exchange. As a result, the selected
limit price is broadly distributed according to a power-law, and
it isn't randomly drawn from the distribution, but has a strong
correlation with the size of cumulative unexecuted limit orders on
the price. It was also found by the price scale transformation of
the conditional probability that the limit price, on the
coarse-grained price scale, tends to gather around the price which
has a large size of cumulative unexecuted limit orders.

\section{Rebuild Order Book and our data}
\label{data}

We analyze the Rebuild Order Book of SETS. In this section, we
give a brief description of the Rebuild Order Book and our data.
SETS is fully automated order book trading service of London Stock
Exchange \cite{lse}. The order book holds details of all orders. A
coming new order to sell or buy will be fully or partially
executed against existing orders on the order book, if both
requirements agree. The unexecuted portion of the order will be
stored on the order book. The Rebuild Order Book is composed of 3
files. Order detail file contains details of new orders, their
prices, sizes and so on. Order history file contains a history of
each order and the method by which it is removed, that is,
deletion, expiry, partial match and full match. Trade report
contains details of every trade, that is, the price, the size and
so on. Merge of these files and sort of records by the time stamp
enable us to pursue the occurrence and the change of each order.

We study the Rebuild Order Book of the 6 months since July to
December in 2004. Our data contains all orders and transactions of
the selected 13 stocks listed on SETS. The selected stocks are
most actively traded in the period, which are the stocks of top 13
bargains in July 2004. There are millions of limit orders,
cancellations and executions in our data. The stocks we analyzed
are Astrazeneca (AZN), Barclays (BARCS), BT Group (BT), BP. (BP),
Diageo (DGE), Glaxosmithkline (GSK), HBOS (HBOS), HSBA HLDGS
(HSBA), Lloyds Tsb Group (LLOY), Shell Tranport \& Trading Co.
(SHEL), Tesco (TSCO), Royal Bank Scot (RBS) and Vodafone Group
(VOD), belonging to various industries, that is, bank, beverages,
oil \& Gas, telecommunication services and so on.

\section{Results of data analysis and discussions}
\label{results}

The purpose of this paper is to examine the correlation of the
limit price with the condition of order book at the time when a
limit order comes. In this section, we report some results of
analysis on the data described in the previous section.

\begin{figure}
\centering
\includegraphics[width=8cm]{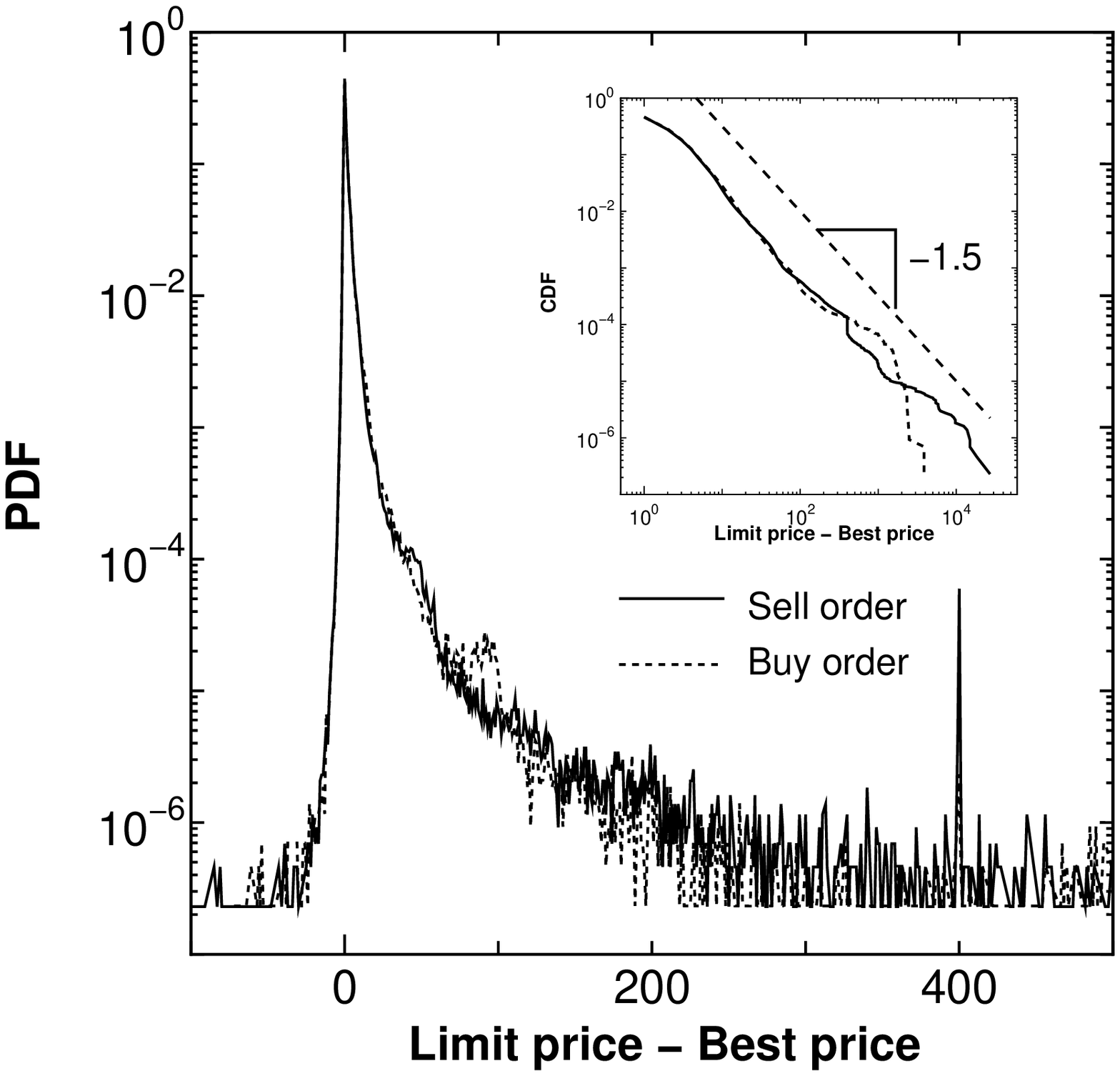}
\caption{Probability distribution function of the relative limit
price. The unit of price is the tick size of each stock. The inset
is the log-log plot of the cumulative distribution function for
limit orders placed on the book. The dashed-line represents the
power-law with the exponent -1.5. } \label{fig1}
\end{figure}

First of all, we show that the coming limit price is broadly
distributed around the best price at the time when the order
comes. Figure \ref{fig1} is the semi-log plot of the probability
distribution function of relative limit price: $price-ask$ (for
sell limit orders), $bid-price$ (for buy limit orders). The unit
of price is the tick size of each stock. The inset is the log-log
plot of the cumulative distribution function for limit orders
placed on the book. The exponent of the power-law tail of the
relative limit price for orders placed on the book is near -1.5,
and the value is consistent with the previous work of this kind
\cite{zovko}.

\begin{figure}
\includegraphics[width=8cm]{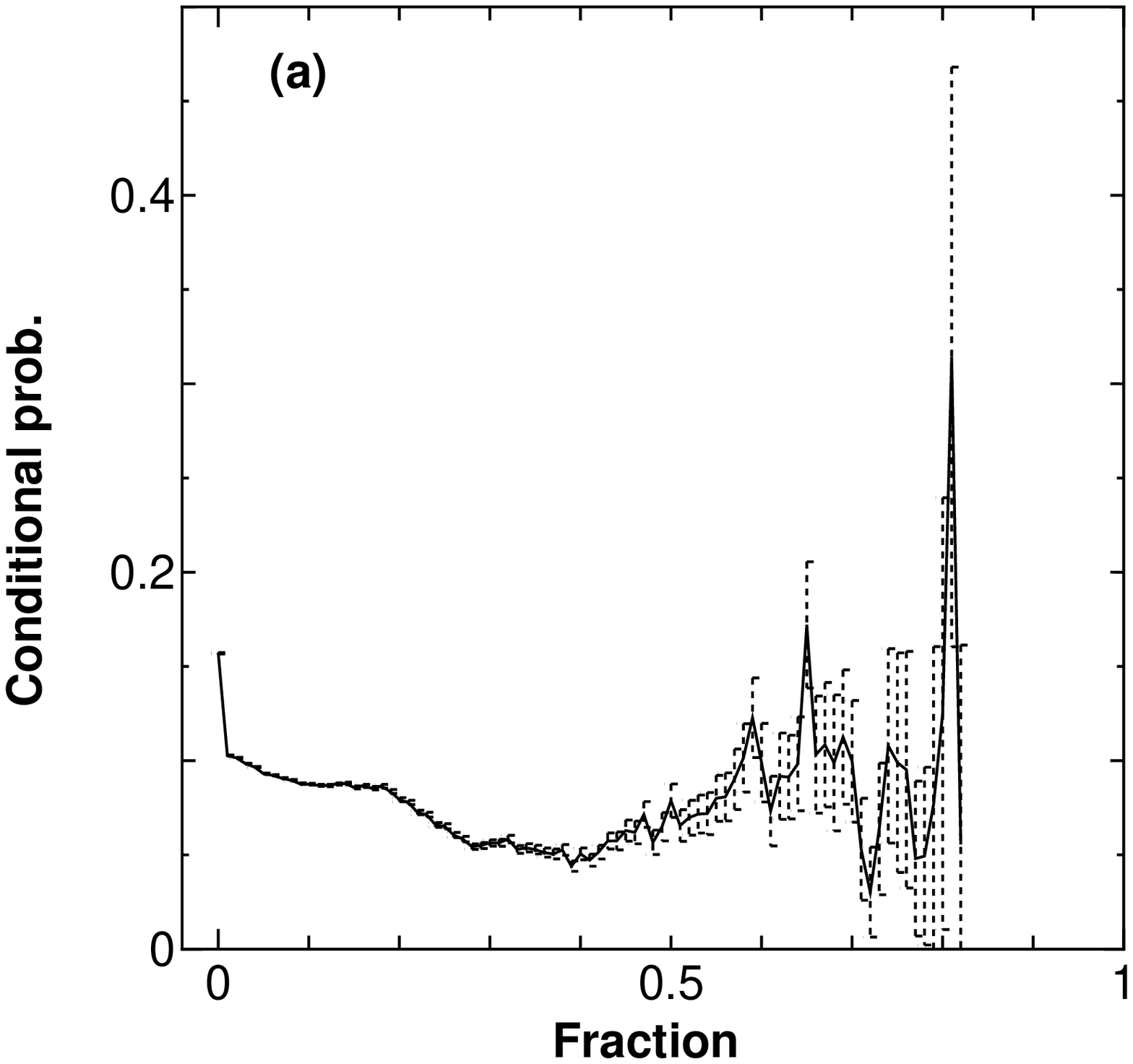}
\includegraphics[width=8cm]{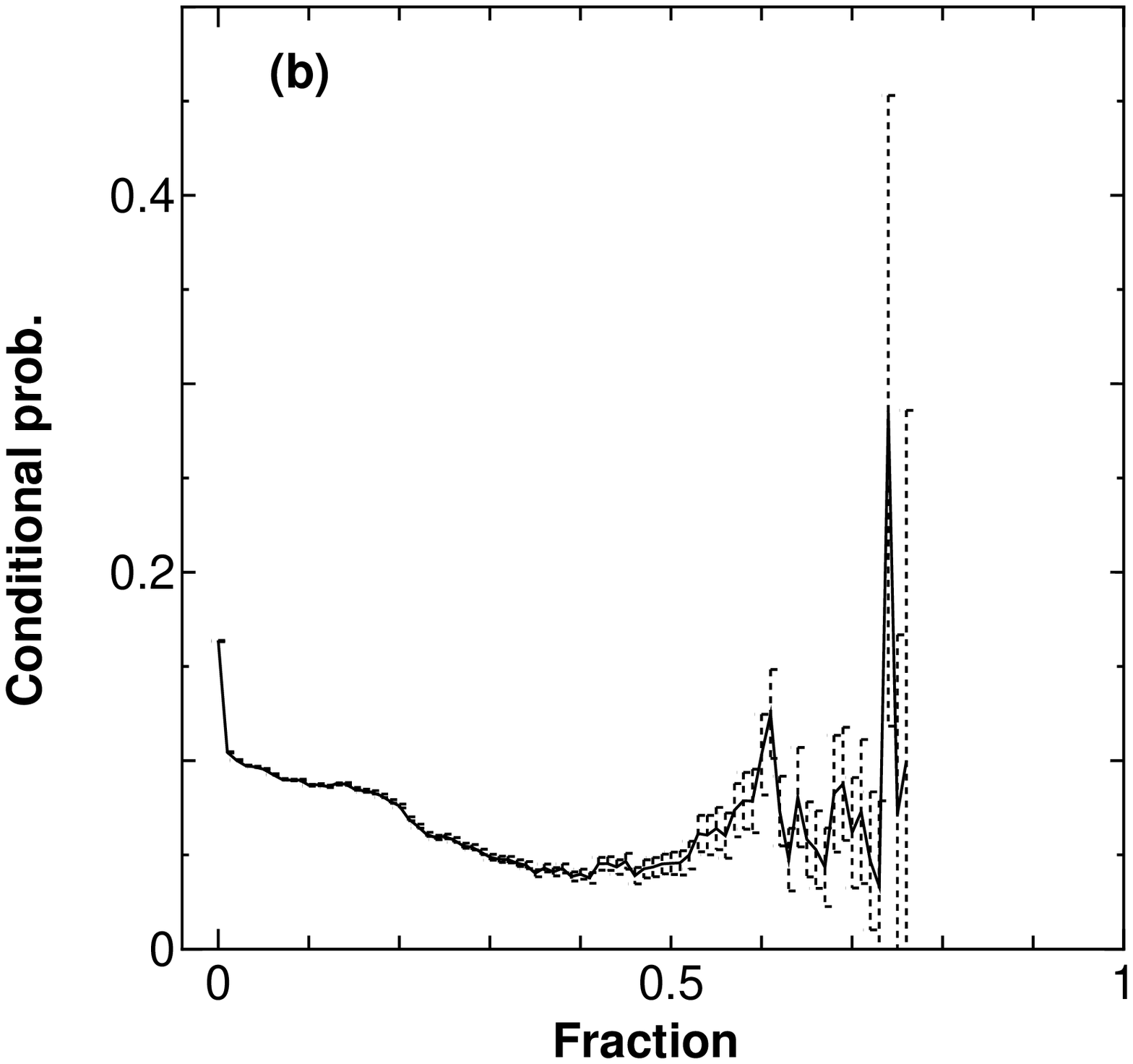}
\caption{Conditional probability that a coming limit order is
placed at the price holding a given fraction $f$. The error bar
represents 95 \% confidential interval. (a)Sell order. (b)Buy
order.} \label{fig2}
\end{figure}

Second, we examine the correlation of the selection of limit price
with the cumulative size of unexecuted limit orders at each price.
Let us consider a situation of the order book. There may be
several prices at which unexecuted limit orders are stored. Each
price may hold a fraction of whole unexecuted orders. Then, which
price will be selected as a coming limit price? Figure \ref{fig2}
shows the conditional probability $P_f$ that a coming limit order
is placed at the price holding a given fraction $f$ of all limit
orders. In both sides of order, The conditional probability $P_f$
decreases with fraction $f$ while $f$ is smaller than about 0.4,
and increases with $f$ while $f$ is smaller than about 0.6. The
cumulative size of unexecuted limit orders at each price
influences the probability of the selection of the price. However,
the conditional probability is not a monotonically increasing
function of $f$.

\begin{figure}
\includegraphics[width=8cm]{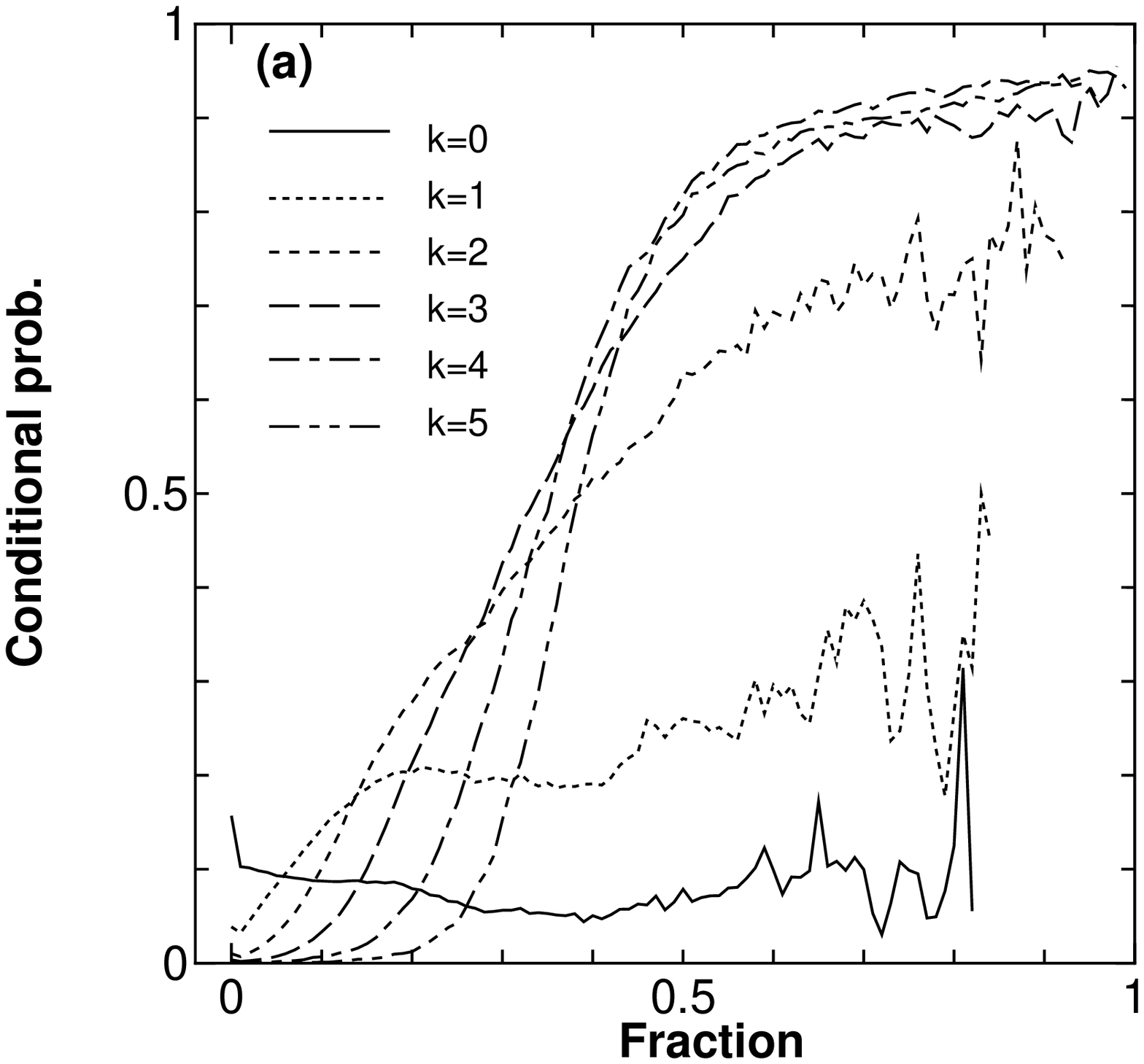}
\includegraphics[width=8cm]{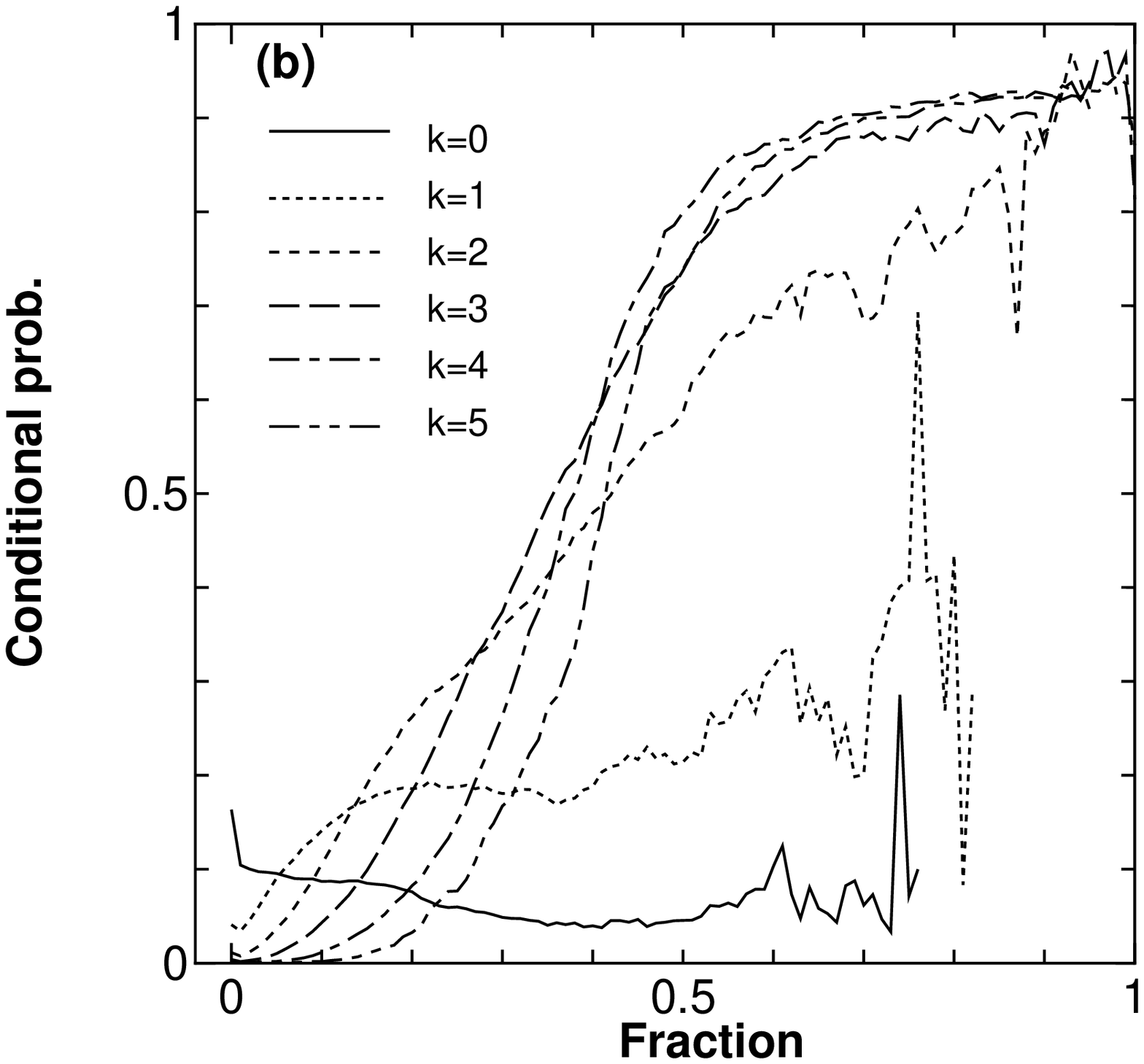}
\caption{Conditional probability on different price scales ($k=0$
to $5$). The parameter k is explained in text. (a)Sell order.
(b)Buy order.} \label{fig3}
\end{figure}

Next, we perform the scale transformation of price, and
renormalize the conditional probability. We group the relative
price $2i$ and $2i+1$ into the group $i$, where the relative price
$i=...,-2,-1,0,1,2,...$ is defined by the equation  $i=price -
ask$ and $i=bid - price$ for sell and buy orders respectively. On
the coarse-grained price scale, we call the group $0$ the best
price (ask/bid), and the group $i$ the relative price $i$ again.
Repeating this procedure k-times, we have the scale transformation
of the relative price, $i^{(k)}=2^k i$. When the stochastic
variable $F^k_i$ denote the cumulative size of limit order at the
price $i^{(k)}$ and the binary stochastic variable $S^k_i=0,1$ the
selection of the price $i^{(k)}$, the renormalization of the
conditional probability for each price is defined by the equation:
\begin{eqnarray}
P(S^k_i=1|F^k_i=f)=\frac{P(S^k_i=1,F^k_i=f)}{P(F^k_i=f)}\nonumber\\
=\frac{\sum\limits_{f^{k-1}_{2i}+f^{k-1}_{2i+1}=f}P(S^{k-1}_{2i}=1
\cup S^{k-1}_{2i+1}=1,F^{k-1}_{2i}=f^{k-1}_{2i} \cap
F^{k-1}_{2i+1}=f^{k-1}_{2i+1})}{\sum\limits_{f^{k-1}_{2i}+f^{k-1}_{2i+1}=f}P(F^{k-1}_{2i}=f^{k-1}_{2i}
\cap F^{k-1}_{2i+1}=f^{k-1}_{2i+1})}.
\label{renormalization}
\end{eqnarray}
The conditional probability $P^k_f$, on the coarse-grained price
scale, is expressed by the weighted average $P^k_f=\sum\limits_i
P(F^k_i=f|\cup_j F^k_j=f)P(S^k_i=1|F^k_i=f)$. The results for
$k=0$ to $5$ ($k=0$ means the original price scale) are shown in
Fig. \ref{fig3}. On the coarse-grained price scales $k \ge 1$, the
conditional probabilities are monotonically increasing function of
$f$. It means that the limit price, on the coarse-grained price
scale, tends to gather around the price which has a large size of
cumulative unexecuted limit orders.

\begin{figure}
\includegraphics[width=8cm]{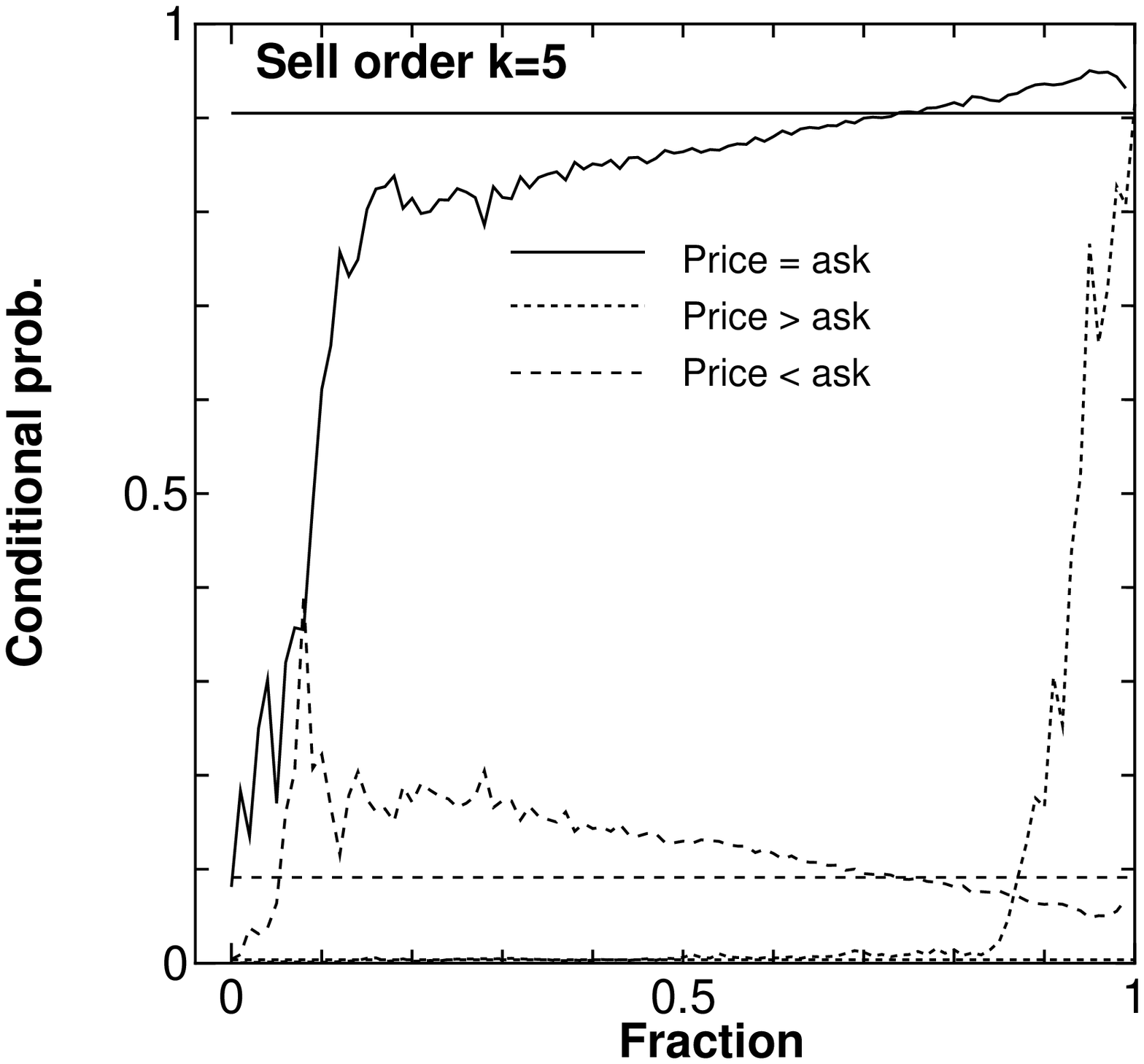}
\includegraphics[width=8cm]{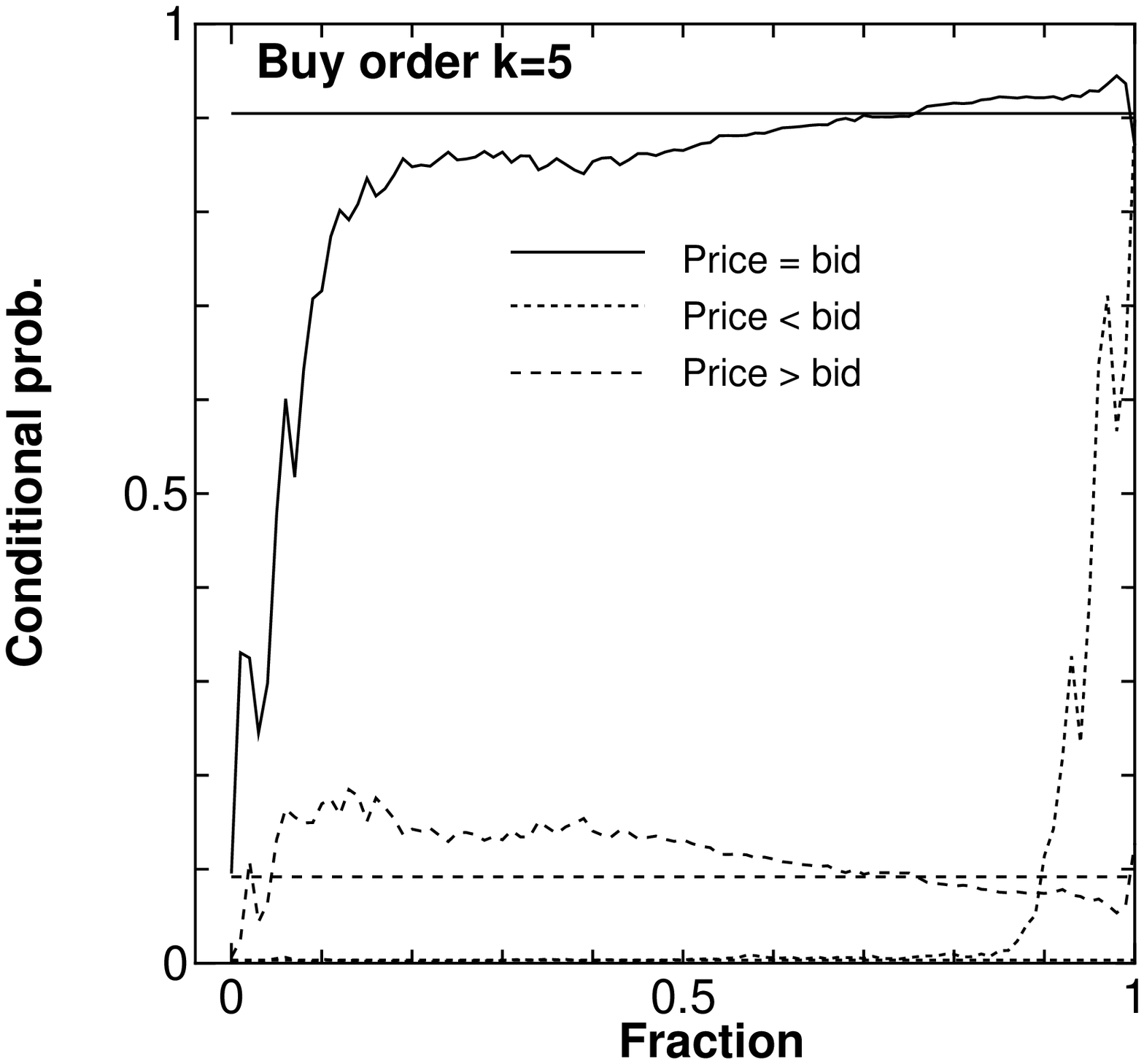}
\caption{Conditional probability on a coarse-grained price scale
($k=5$). Prices are divided into three groups, the best price,
prices on the book and prices in the spread. The horizontal axis
for the prices in the spread is the fraction $f$ of limit orders
held at the best price. The horizontal line for each group
represents the unconditional probability that traders select the
group.} \label{fig4}
\end{figure}

Finally, we study the dependency of the conditional probability on
the relative price $i^{(k)}$. We divide the whole prices into
three groups, $i^{(k)}=0$ (the best price), $i^{(k)}>0$ (prices on
the book) and $i^{(k)}<0$ (prices in the spread), and derive the
corresponding conditional probabilities $P(S^k_0=1|F^k_0=f)$,
$P(\cup_{i>0}S^k_i=1|\sum_{i>0}F^k_i=f)$ and
$P(\cup_{i<0}S^k_i=1|F^k_0=f)$. The conditional probability
$P(\cup_{i>0}S^k_i=1|\sum_{i>0}F^k_i=f)$ is derived from the
equation $P(S^k_0=1|F^k_0=1-f)+
P(\cup_{i>0}S^k_i=1|\sum_{i>0}F^k_i=f)+P(\cup_{i<0}S^k_i=1|F^k_0=1-f)=1$.
The results for sell and buy order on the price scale $i^{(5)}=2^5
i$ ($k=5$) are shown in Fig. \ref{fig4}. The unconditional
probability that traders select the group is also shown by the
horizontal line. On this coarse-grained price scale, each price
contains 32 original price levels. In the region of $f$ close to
1, the conditional probability that traders select the prices
separate from the best price by over 32 original price levels is
hundreds times as large as the unconditional probability, which is
tiny. In the region $f<0.07$, the inequality
$P(\cup_{i\le0}S^k_i=1|F^k_0=f)<P(\cup_{i>0}S^k_i=1|F^k_0=f)$
holds, which means the state such that the stored limit orders are
sparse within the distance of 31 original price levels from the
best price is stable. As is shown by the simulation in the paper
\cite{maskawa2006}, such unequal attractive power of prices has
amplified the fluctuation of gaps between occupied price levels,
and cause the power-law fluctuation of price changes and the long
memory of the volatility. This result agrees with the statement by
Farmer et al. in the paper \cite{farmer2004}. They has argued that
large returns are not caused by large orders, while the large gaps
between the occupied price levels on the order book lead to large
price changes in each transaction.

\section{Conclusions}
\label{conclusions}

We have analyzed the Rebuild Order Book of SETS, which is fully
automated order book trading service of London Stock Exchange in
order to examine the correlation of the limit price with the
condition of order book at the time when a limit order comes.

We have found that there is a obvious correlation. Especially on
the coarse-grained price scale, our data clearly shows that the
limit price tends to gather around the price at which a large size
of unexecuted limit orders has been stored. Such unequal
attractive power of prices is a promising candidate for the origin
of the fat tail of the price fluctuation and the long memory of
the volatility.

This work was supported by the Japan Society for the Promotion of
Science under the Grant-in-Aid, No. 15201038.



\newpage

\newpage

\end{document}